\documentclass[prb,twocolumn,longbibliography,
superscriptaddress,
 amsmath,amssymb,
 aps,floatfix
]{revtex4-2}
\usepackage{gensymb}
\usepackage{graphicx}
\usepackage{dcolumn}
\usepackage{bm}
\usepackage{multirow}
\usepackage{textcomp, gensymb}
\usepackage{
}

\DeclareUnicodeCharacter{2212}{-}
\newcommand{\nps}{NiPS$_3$}
\newcommand{\evbm}{$E-E_{VBM}$}
\newcommand{\vbm}{$E_{VBM}$}
\usepackage{xcolor}

\begin{document}

\title{Revealing the Electronic Structure of van der Waals Antiferromagnetic NiPS\textsubscript{3} through Synchrotron-Based $\mu$-ARPES and Alkali Metal Dosing}

\author{Yifeng Cao}
\thanks{These authors contributed equally to this work.}
\affiliation{Department of Physics, Boston University, Boston, Massachusetts, 02215, USA.}
\affiliation{Advanced Light Source, Lawrence Berkeley National Laboratory, Berkeley, California, 94720, USA.}

\author{Qishuo Tan}
\thanks{These authors contributed equally to this work.}
\affiliation{Department of Chemistry, Boston University, Boston, Massachusetts, 02215, USA.}

\author{Yucheng Guo}
\thanks{These authors contributed equally to this work.}
\affiliation{Department of Physics and Astronomy, Rice University, Houston, Texas, 77005, USA.}
\affiliation{Advanced Light Source, Lawrence Berkeley National Laboratory, Berkeley, California, 94720, USA.}

\author{Clóvis Guerim Vieira}
\affiliation{Departamento de Física, Universidade Federal de Ouro Preto, Ouro Preto, CEP 35400-000, Brazil.}

\author{M\'ario S. C. Mazzoni}
\affiliation{Departamento de Física, Universidade Federal de
Minas Gerais, Belo Horizonte, CEP 31270-901, Brazil.}

\author{Jude Laverock}
\affiliation{School of Chemistry, University of Bristol, Bristol, BS8 1QU, UK.}

\author{Nicholas Russo}
\affiliation{Department of Physics, Boston University, Boston, Massachusetts, 02215, USA.}
\affiliation{Advanced Light Source, Lawrence Berkeley National Laboratory, Berkeley, California, 94720, USA.}

\author{Hongze Gao}
\affiliation{Department of Chemistry, Boston University, Boston, Massachusetts, 02215, USA.}

\author{Chris Jozwiak}
\affiliation{Advanced Light Source, Lawrence Berkeley National Laboratory, Berkeley, California, 94720, USA.}

\author{Aaron Bostwick}
\affiliation{Advanced Light Source, Lawrence Berkeley National Laboratory, Berkeley, California, 94720, USA.}

\author{Eli Rotenberg}
\affiliation{Advanced Light Source, Lawrence Berkeley National Laboratory, Berkeley, California, 94720, USA.}

\author{Jinghua Guo}
\affiliation{Advanced Light Source, Lawrence Berkeley National Laboratory, Berkeley, California, 94720, USA.}

\author{Ming Yi}
\affiliation{Department of Physics and Astronomy, Rice University, Houston, Texas, 77005, USA.}

\author{Matheus J. S. Matos}
\email{matheusmatos@ufop.edu.br}
\affiliation{Departamento de Física, Universidade Federal de Ouro Preto, Ouro Preto, CEP 35400-000, Brazil.}

\author{Xi Ling}
\email{xiling@bu.edu}
\affiliation{Department of Chemistry, Boston University, Boston, Massachusetts, 02215, USA.}
\affiliation{Division of Materials Science and Engineering, Boston University, Boston, Massachusetts, 02215, USA. }
\affiliation{The Photonics Center, Boston University, Boston, Massachusetts, 02215, USA.}

\author{Kevin E. Smith}
\email{ksmith@bu.edu}
\affiliation{Department of Physics, Boston University, Boston, Massachusetts, 02215, USA.}
\affiliation{Department of Chemistry, Boston University, Boston, Massachusetts, 02215, USA.}
\affiliation{Division of Materials Science and Engineering, Boston University, Boston, Massachusetts, 02215, USA. }


\begin{abstract}
Antiferromagnetic NiPS\textsubscript{3} has recently emerged as a quantum material of considerable interest, thanks to the discovery of multiple new couplings involving electrons, spins, orbitals, phonons, and magnons. However, controversies and open questions persist concerning the fundamental origins of these couplings. A critical piece of information required to advance the understanding is the precise electronic band structure of NiPS\textsubscript{3}. Angle-resolved photoemission spectroscopy (ARPES), combined with alkali metal dosing (AMD), can enable us to directly observe the subtle and novel electronic states that appear around the Fermi surface, offering valuable insights into the intriguing quantum properties and interplays of the examined material. Here, we present a comprehensive characterization and analysis of the band structure of van der Waals layered antiferromagnet NiPS\textsubscript{3}, leveraging state-of-the-art $\mu$-ARPES measurements supported by density functional theory (DFT) calculations. Theoretical DFT results identify the orbital contributions to the observed bands, providing a precise understanding of the experimental ARPES data. Crucially, AMD enables the observation of conduction band and defect-related states above the valence band maximum in NiPS\textsubscript{3}. Furthermore, temperature dependent ARPES results across the Néel transition temperature of NiPS\textsubscript{3} reveal that the paramagnetic and antiferromagnetic phases have nearly identical band structures, underlining the highly localized character of Ni $d$ states. These findings substantially deepen our understanding of the electronic properties of NiPS\textsubscript{3} and lay a vital foundation for exploring the intriguing quantum phenomena it exhibits.

\end{abstract}

\maketitle

\section{Introduction}

Quantum materials have become a key focus in condensed matter physics and materials science, offering the promise of transformative technologies and deeper insights into fundamental physics. They host exotic phenomena—such as topological phases, unconventional superconductivity, and complex magnetic orders—driven by strong electronic correlations\cite{basov2017towards, keimer2017physics}. Exploring these materials paves the way for breakthroughs in quantum computing\cite{de2021materials}, spintronics\cite{han2016perspectives}, and next-generation electronic and photonic devices\cite{bauer2020quantum, carusotto2020photonic}. Among the wide variety of quantum materials, NiPS\textsubscript{3} has garnered particular attention for its noteworthy properties in electronics\cite{kim2018charge, tan2022charge, cao2024spectral}, spintronics\cite{basnet2021highly, scheie2023spin}, magnetism\cite{wildes2015magnetic, kim2019suppression, he2024magnetically, Tan2024}, and optics\cite{liu2019nips, wang2022electronic}. One of its most striking features is the ultrasharp photoluminescence (PL) emission\cite{kang2020coherent, wang2021spin, hwangbo2021highly} observed below the Néel transition temperature ($T_N$). This PL emission not only has strict spin-induced linear polarization\cite{wang2021spin} that enables a new optical probe of in-plane antiferromagnetic (AFM) order\cite{Tan2024}, but also shows strong coupling to lattice vibrations\cite{hwangbo2021highly}. However, its origin is still controversial with several potential interpretations such as coherent many-body exciton\cite{kang2020coherent}, band-edge emission\cite{wang2021spin, ho2021band}, defect emission\cite{Xiaoqin}, phonon-bound exciton\cite{hwangbo2021highly} or Hund's exciton\cite{he2024magnetically}. The debate stems from the complexity of the correlated Ni \(3d\) electrons, the subtle interplay between magnetism and lattice vibrations, and the possibility that multiple mechanisms could contribute simultaneously to the observed ultra-sharp PL. Resolving its origin necessitates a full understanding of the electronic band structure.

Angle-resolved photoemission spectroscopy (ARPES) allows for the direct measurement of the electronic structure of the occupied valence band (see Fig.~\ref{fig:1}\textbf{c}). With alkali metal dosing (AMD), charge will transfer from the alkali metal to the surface of the material, occupying the conduction band \cite{muscat1986coverage, aruga1989alkali, diehl1997current}. This makes AMD a widely adopted strategy to probe the unoccupied conduction bands using ARPES \cite{ohta2006controlling, zhang2014direct, smolenski2024large}. To thoroughly explore the comprehensive electronic structure of NiPS\textsubscript{3}, the ARPES method is indispensable. Recently, several studies have utilized lab-based ARPES to investigate the electronic structure of NiPS\textsubscript{3} single crystals\cite{yan2023electronic, klaproth2023origin}, as well as other transition metal phosphorus trichalcogenides, including MnPS\textsubscript{3}\cite{strasdas2023electronic}, CoPS\textsubscript{3}\cite{voloshina2023arpes} and FePS\textsubscript{3}\cite{yan2023electronic}. However, although a small number of bands were observed, the majority of bands were either fuzzy or even invisible due to the limited resolution of lab-based ARPES and the severe charging effect induced by the relatively large band gap of NiPS\textsubscript{3}\cite{brec1979physical,jenjeti2016alternate,kim2018charge, ho2021band}. Moreover, density functional theory (DFT) calculations are commonly used to elucidate ARPES experimental results\cite{yan2023electronic}; however, the above issues impose limitations to the agreement between theory and the ARPES data. Therefore, a high resolution ARPES that can be synergistically analyzed with the calculated electron band structure is highly desirable for understanding the emerging couplings in the material and resolving ongoing controversies in the field.

Here, we present a comprehensive synchrotron-based $\mu$-ARPES investigation of NiPS\textsubscript{3}. By systematically tuning photon energies and polarizations of the beam, we resolve the band structure from the valence band maximum (VBM) down to $E-E_{VBM}$ = -7.3 eV with high clarity. Theoretical calculations using DFT are in agreement with the experimental data, facilitating the analysis of contributions from Ni, S and P orbitals to the observed bands. In addition, we employ AMD at low temperature to access states above the VBM, thereby revealing both conduction band features and defect-related states. Further DFT modeling of the AMD process provides insights on how the band structures change with potassium doping effect. Moreover, temperature-dependent ARPES measurements across $T_N$ show that both paramagnetic and antiferromagnetic phases share nearly identical band structures, which is a notable contrast to the behavior observed in MnPS\textsubscript{3}\cite{strasdas2023electronic}, reinforcing the picture of highly localized Ni $d$ states in NiPS\textsubscript{3}. 

\section{Methods}

\begin{figure*}[htb]
\centering
\includegraphics[width=15.7cm]{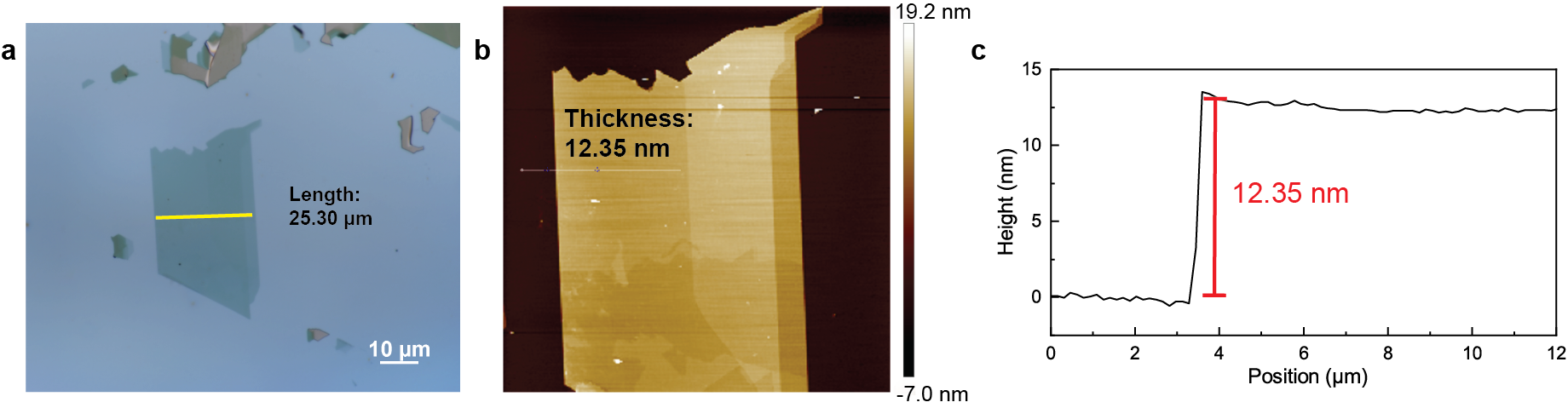}
\caption{\textbf{NiPS\textsubscript{3} sample characterization.} (a) Optical microscope image of a nanometer-thin NiPS\textsubscript{3} flake, with the yellow solid line indicating the length of the target flake. (b) Atomic force microscope image of the same flake. (c) Height profile showing the measured thickness.}
\label{fig:sample}
\end{figure*}

\textbf{NiPS\textsubscript{3} sample preparation.} Pure elements, in a stoichiometric ratio of Ni:P:S $=$ 1:1:3 (2 g in total), along with 40 mg iodine as transport agent, were enclosed in quartz ampules under vacuum of 1$\times$10\textsuperscript{$-$4} Torr. Subsequently, the ampules underwent heating in a two-zone furnace with the temperature range of 600 $-$ 650 $^{\circ}$C for one week, followed by cooling to room temperature. Bulk crystals were harvested from the lower temperature zone of the ampules. Thin flakes were acquired through mechanical exfoliation on boron-doped Si substrates (coated with 1 nm/5 nm titanium/gold film to enhance the flake adhesion using scotch-tape, and their thicknesses were determined by atomic force microscopy (Bruker Dimension 3000). We selected target samples with a thickness of approximately 10-50 nm and an area with a uniform thickness greater than 20×20 µm (see Fig.~\ref{fig:sample}).

\begin{figure*}[htb]
\centering
\includegraphics[width=15.7cm]{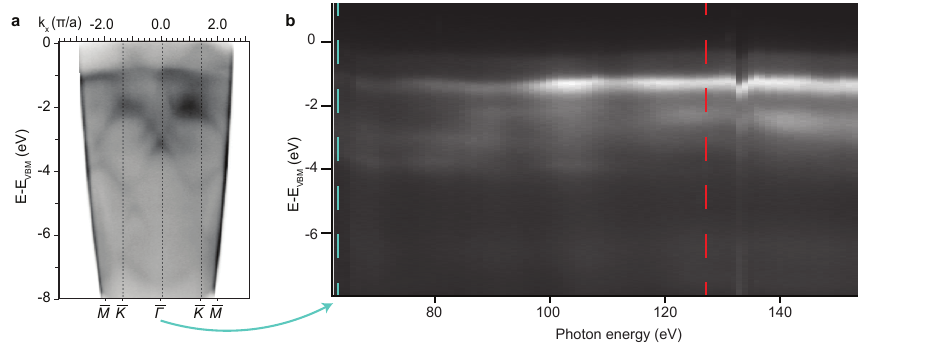}
\caption{\textbf{Photon energy dependent ARPES results.} (a) Measured dispersion along the $\overline{MK\Gamma KM}$ direction at 6 K with LH beam at $h\nu$ = 63 eV, as indicated by the blue dashed line in panel (b). (b) Photon energy dependent measurement at $\overline{\Gamma }$ point with LH beam at 6 K. The red dashed line in right panel indicates the photon energy at $h\nu$ = 126 eV, which has been applied in the current work.}
\label{fig:hv}
\end{figure*}

\textbf{ARPES setup.} We probed the NiPS\textsubscript{3} with $\mu$-ARPES at Advanced Light Source beamline 7.0.2 at Lawrence Berkeley National Laboratory. This beamline is equipped with an R4000 spectrometer with deflectors, allowing data collection across a full Brillouin zone without the need to move the sample. The $\mu$-ARPES has spot size less than 10 µm and energy resolution down to 16 meV. The NiPS\textsubscript{3} nanometer-thin flakes on boron-doped Si substrates were mounted on Cu pucks secured with Ti clamps. Epotek H20E silver epoxy was also applied between the surface of the Si substrate and the Cu puck to increase electronic conductivity. Prior to $\mu$-ARPES measurements, the samples were transferred to the preparation chamber and annealed at 350°C for 12 hours to remove the surface contamination. The photon energy in $\mu$-ARPES was tunable and we covered the range from 63 eV to 153 eV (see Fig.~\ref{fig:hv}\textbf{b}). The experiment revealed that the photon energy of 126 eV with linear horizontal polarization (see Fig.~\ref{fig:1}\textbf{c}) provides the clearest band structure for NiPS\textsubscript{3}. AMD was performed by depositing potassium onto the surface of a NiPS\textsubscript{3} flake at 7 K. A well-outgassed potassium getter source (from SAES Industrial), operated at a current of 7 A, was used for dosing, with each round lasting for 1 minute. To obtain the spectra across the $T_N$ (155 K) of NiPS\textsubscript{3}, the temperature range of our experiments spans from 6 K to 182 K.

\textbf{Computational methods} \textit{Ab initio} calculations were performed using density functional theory with on-site Coulomb correlations (DFT+U) to investigate the band structure of NiPS\textsubscript{3}\cite{hohenberg1964inhomogeneous, PhysRev.140.A1133, Dudarev1998}. These calculations were carried out with the VASP code, employing the projector-augmented wave (PAW) method\cite{Kresse1993, Kresse1996, Blchl1994, Kresse1999, PhysRevB.54.11169}. To account for the on-site Coulomb repulsion between the 3\textit{d} electrons of Ni ions, we applied the GGA+U formalism proposed by Dudarev~\cite{Dudarev1998}, with effective Hubbard U parameters ranging from 1.0 eV to 4.0 eV, in conjunction with the Perdew-Burke-Ernzerhof (PBE) exchange-correlation functional \cite{Perdew1996}. Band structure plotting was conducted utilizing band unfolding techniques as implemented in the VASPKIT software~\cite{Wang2021}, facilitating direct comparison with ARPES experimental data. We compare DFT+U electronic band positions with ARPES data to determine the optimal U parameter for pristine and K-adsorbed NiPS\textsubscript{3}. The plane-wave basis set employed in the calculations utilized an energy cutoff of 515 eV and the criteria for energy convergence was set to $10^{-5}$ eV. Geometries were optimized until the maximum atomic force was below 10 meV/\r{A}. We simulate the zigzag antiferromagnetic configuration in NiPS\textsubscript{3} employing a 2×2×1 supercell for bulk calculations and 2×2 supercell for monolayer calculations. To achieve geometric relaxation of the structures, a Gamma-centered k-grid of 6×6×1 was employed. To accurately describe Van der Waals interactions between layers, we use the DFT-D3 method of Grimme with zero-damping~\cite{10.1063/1.3382344}. The density of states (DOS) analysis was performed using the VASPKIT software~\cite{Wang2021}. To simulate theoretical constant energy maps, the WannierTools package~\cite{Wu2018} was utilized (for more details of the computation method, see Section. S2 in the Supplemental Material\cite{SM}. The Supplemental Material also contains Refs.\cite{Mostofi2014,abbate1993soft,yang2021situ,wang2018new,wang2001edge}).

\section{Results}

\begin{figure*}[htb]
\centering
\includegraphics[width=15.7cm]{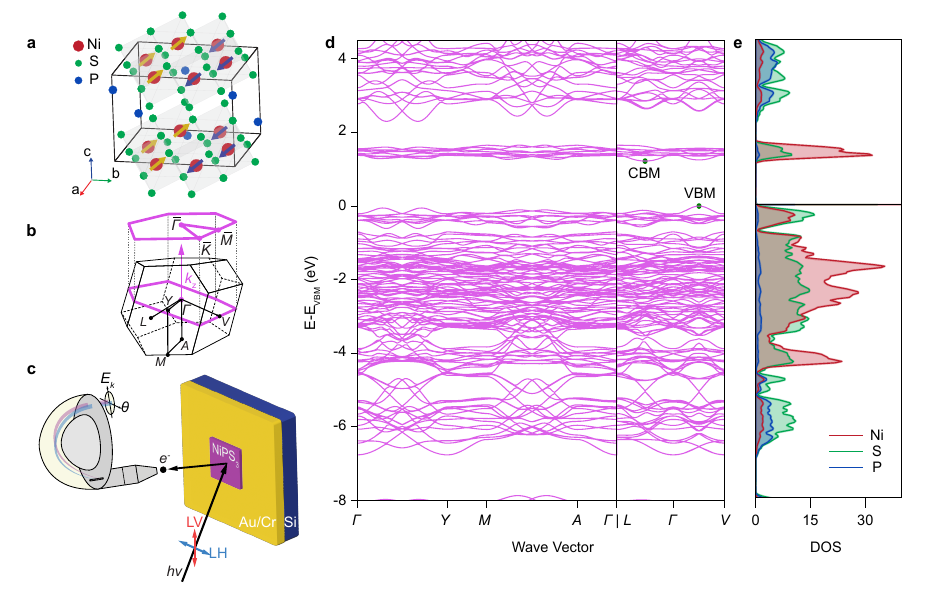}
\caption{\textbf{Crystal structure and calculated electronic band structure of NiPS\textsubscript{3}.} (a) Crystal and magnetic structure of NiPS\textsubscript{3}. Yellow and blue arrows stand for spin-up and spin-down in its antiferromagnetic state. (b) Three-dimensional Brillouin zone with an arrow indicating the direction of $k_z$. Hexagon on the top is the projection of the surface Brillouin zone. (c) Stratified arrangement of the NiPS\textsubscript{3} flakes on Au/Cr coated Si substrate and beam geometry with LH and LV standing for linear horizontal and linear vertical polarizations, respectively. (d) DFT calculated unfolded electronic structure of bulk NiPS\textsubscript{3} along all high-symmetry points and (e) the integrated density of states (DOS) with Ni, S, and P contributions labeled in red, green, and blue, respectively. Conduction band minimum (CBM) and valence band maximum (VBM) are labeled with green dots.}
\label{fig:1}
\end{figure*}

The crystalline structure of NiPS\textsubscript{3} is classified within the monoclinic \textit{C}2/\textit{m} space group~\cite{OUVRARD19851181}, and is distinguished by a honeycomb lattice arrangement of Ni\textsuperscript{2+} ions encapsulating [P\textsubscript{2}X\textsubscript{6}]\textsuperscript{4-} bipyramids as can be seen from Fig.~\ref{fig:1}\textbf{a}, and the corresponding Brillouin zone and projected surface Brillouin zone are displayed in Fig.~\ref{fig:1}\textbf{b}. In addition, the magnetic structure of NiPS\textsubscript{3} exhibits an in-plane zigzag-type antiferromagnetic ordering below the $T_N$ at 155 K\cite{le1982magnetic}, and corresponds to the magnetic space group P$_C$2$_1$/m~\cite{PhysRevB.92.224408,PhysRevB.106.174422}.  Figure~\ref{fig:1}\textbf{c} shows a setup of the synchrotron-based $\mu$-ARPES, where samples are placed in a vacuum chamber, and beams with linear horizontal (LH) and linear vertical (LV) polarizations are used as incident lights. In addition, the theoretical calculated electronic band structure and the corresponding density of states (DOS) are shown in Fig.~\ref{fig:1}\textbf{d} and \textbf{e}, respectively. To obtain a clear ARPES signal from the insulating NiPS\textsubscript{3}, we exfoliate the sample into nanometer-thin flakes to reduce the charging effect. We find that flakes that are too thin (e.g. below 5 layers) will have insufficient sample stiffness, resulting in  poor surface flatness, making it difficult to observe dispersive states. Empirically, it is possible to conclude that samples with a thickness of 10-50 nm yield the best ARPES signals after conducting tests on several samples with various thicknesses (see Fig.~\ref{fig:sample}).

\begin{figure*}[htb]
\centering
\includegraphics[width=15.7cm]{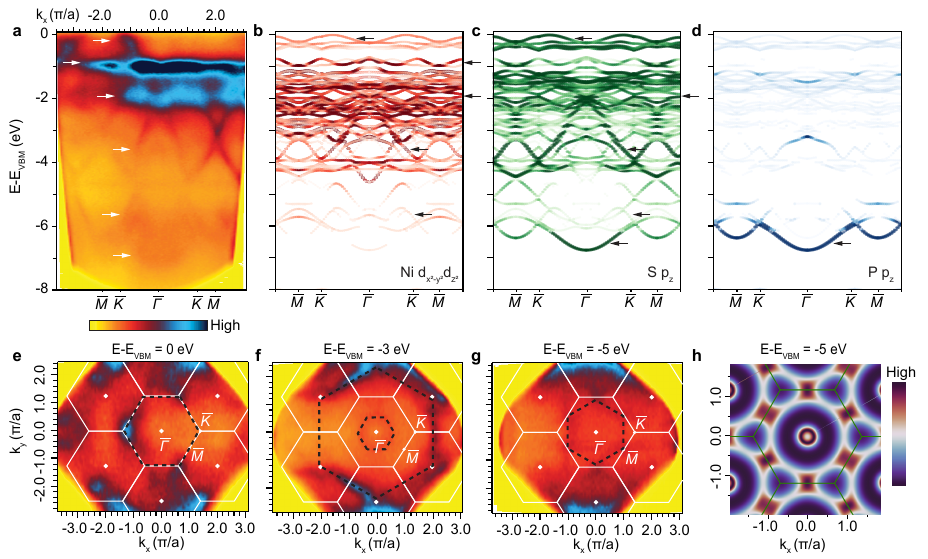}
\caption{\textbf{Observation of the electronic band structure of NiPS\textsubscript{3}.}  (a) Experimental band dispersion along the $\overline{MK\Gamma KM}$ direction at 6 K with LH beam at $h\nu$ = 126 eV. The white arrows indicate the key features of the observed bands. (b)-(d) Band structure calculated by DFT+U approach (U=2.0 eV). The main contributing atomic orbitals are plotted, including the \( d_{x^2 - y^2} \) and \( d_{z^2} \) of Ni and \( p_z \) of S and P. The electronic states of different elements are distinguished by color: red, green, and blue represent Ni, S, and P orbitals, respectively. The intensity of the color indicates the band weight. The black arrows correspond to the white arrows in (a), indicating the primary contribution of different atomic orbitals to the experimentally obtained bands. (e)-(g) Constant energy maps at $E-E_{VBM}$ = 0 eV, -3 eV, and -5 eV, with surface Brillouin zone boundaries labeled by white solid lines and hexagonal pockets labeled by the black dashed lines. (h) Computed spectral function energy cuts at -5 eV for the non-magnetic cell obtained using WannierTools.}
\label{fig:4}
\end{figure*}

\begin{figure*}[htb]
\centering
\includegraphics[width=15.7cm]{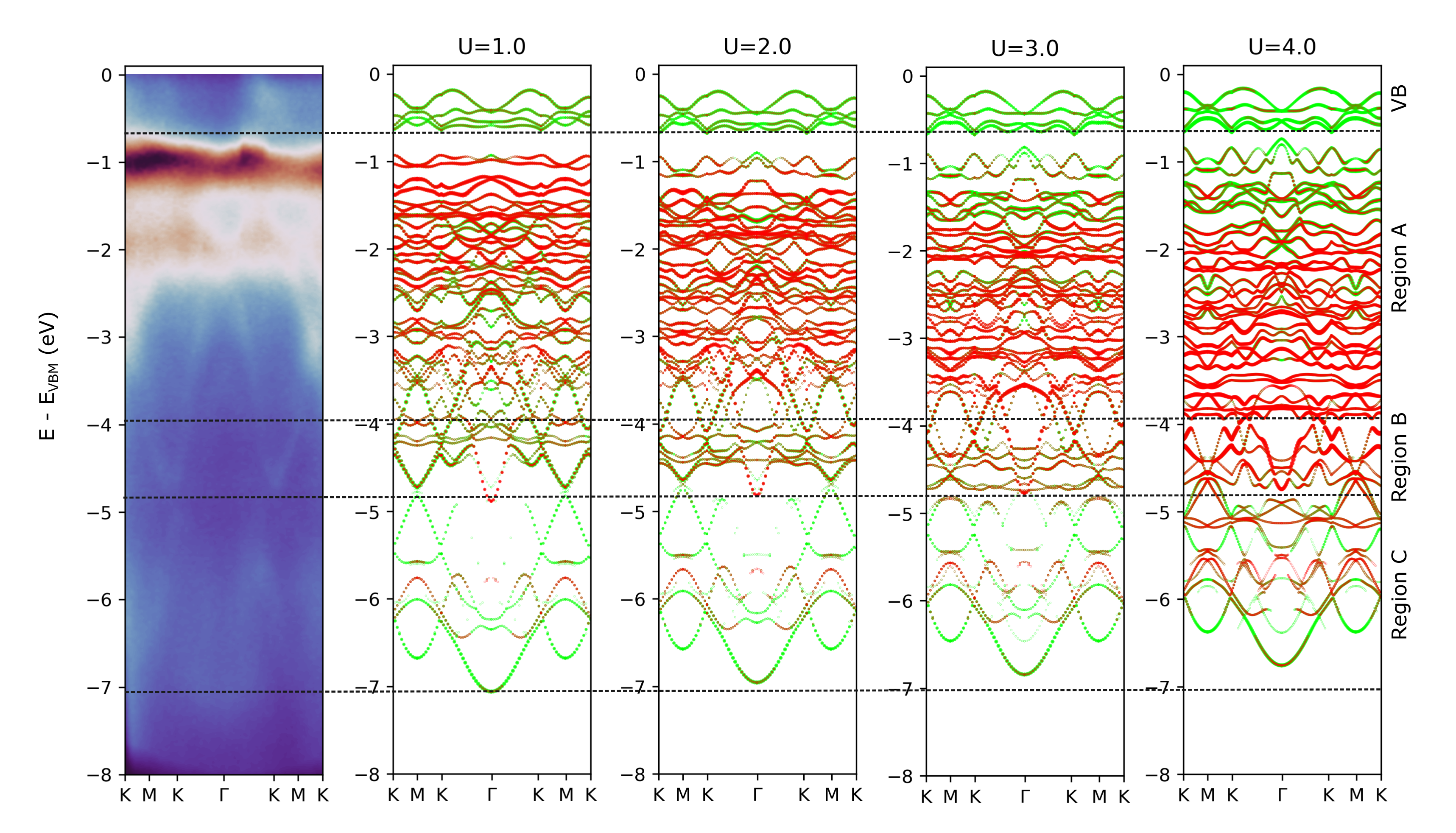}
\caption{\textbf{Comparison between ARPES spectra and calculated bands for different $U$ values in the magnetic bulk geometry.} Ni $d_{z^2}$ orbitals are shown in red and S $p_z$ orbitals in green. For clarity, both the ARPES data and the calculated bands are divided into four distinct regions-VB, Region A, Region B, and Region C-to facilitate direct comparison.
}
\label{fig:U}
\end{figure*}

The detailed band structure of NiPS\textsubscript{3} along the $\overline{MK\Gamma KM}$ direction is presented in Fig.~\ref{fig:4}\textbf{a}, with white arrows indicating the position of the bands. Here, the valence band maximum energy ($E_{VBM}$) is defined by the kink feature in the energy distribution curve (EDC) (see Fig. S1 in the Supplemental Material\cite{SM}). To better interpret our ARPES data, we applied the DFT+U method and determined the optimal value of the Hubbard parameter $U$, as shown in Fig.~\ref{fig:U}. For clarity, both the ARPES spectra and the calculated bands are divided into four regions: VB, Region A, Region B, and Region C. A flat band is observed in the VB region, while several parabolic bands appear in Region C. The experimental ARPES data is particularly clear in Regions A and B, where most of the key features are located. In Region A, two inverted parabolas intersect at approximately $-2.3$ eV, forming a distinct X-shaped structure that is especially prominent at $U = 2.0$ eV and $U = 3.0$ eV. In Region B, V-shaped features appear near the K point in ARPES and are reproduced in the DFT calculations at $U = 1.0$ eV and $U = 2.0$ eV. Based on these comparisons, we conclude that $U = 2.0$ eV provides the best agreement with the experimental data (for further details on tuning the Hubbard parameter, see Section S2.2 in the Supplemental Material\cite{SM}).

To identify the orbital contributions to the measured band structure, we compare the ARPES data with the band structure calculated from DFT+U method, with $U = 2.0$ eV (Fig.~\ref{fig:4}\textbf{b}-\textbf{d}). For simplicity, we only show the atomic orbitals that primarily contribute to the bands indicated by white arrows in Fig.~\ref{fig:4}\textbf{a}, including the \(d_{x^2-y^2}\) and \(d_{z^2}\) orbitals of Ni (red), and the \(p_z\) orbitals of S (green) and P (blue). By comparison, the band closest to \( E_{VBM} \) (around -0.5 eV) is largely derived from Ni and S states, exhibiting a pronounced asymmetry manifested as strong intensity near the  left $\overline{K}$ point. This behavior is also seen in the constant energy map at \( E_{VBM} \) in Fig.~\ref{fig:4}\textbf{e}, where the central hexagonal pocket displays high intensity only at the alternating $\overline{K}$ points. A notably flat band, dominated by Ni $d$ states, appears at \evbm~= -1 eV. Its pronounced flatness supports the picture in which Ni $d$ orbitals form highly localized levels embedded within the broader bands formed by the P\textsubscript{2}S\textsubscript{6} lattice\cite{khumalo1981reflectance}. In the energy range between -2 eV and -6 eV, several additional bands emerge, predominantly arising from the hybridized Ni \( d \) and S \( p \) states. The contribution from P \( p \) states predominantly occurs at deeper energies: a hybridized state of P and S appears at around -7.3 eV relative to \(E_{VBM}\). Figure~\ref{fig:4}\textbf{e}-\textbf{g} present the constant energy maps at various energies, revealing the threefold symmetry of the valence band. This symmetry is reflected in hexagonal pockets of different sizes (indicated by black dashed lines) around the $\overline{\Gamma}$ point at various energies \evbm. Meanwhile, the calculated constant energy map at -5 eV (Fig.~\ref{fig:4}\textbf{h}) shows strong intensity at $\overline{M}$ points and similarly sized hexagonal pockets. Furthermore, calculated constant energy maps also align very well with the experimental results (see Fig. S7 in the Supplemental Material\cite{SM}).

\begin{figure*}[htb]
\centering
\includegraphics[width=15.7cm]{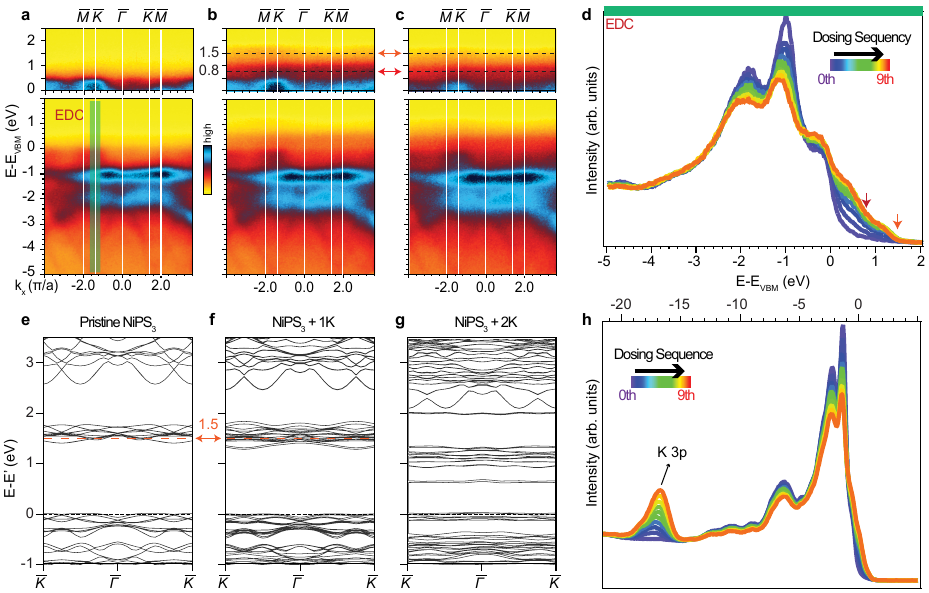}
\caption{\textbf{Conduction band and defect-related state through alkali metal dosing.} (a)-(c) ARPES data with (a) pristine sample before dosing, (b) dosing for 5 rounds, and (c) dosing for 9 rounds along the $\overline{MK\Gamma KM}$ direction. The dosing measurements were taken at 7 K with same LH beam at $h\nu$ = 126 eV. A different intensity scale is applied to the upper panels of (a)–(c) to enhance contrast in the conduction band regions. The orange and red arrows representing the two newly appeared bands after dosing. (d) EDC at the $\overline{K}$ point as indicated by the green line in (a), with color bar shows the dosing sequence. The arrows indicate the onsets of the newly appeared bands in (b). (e)-(g) Calculated structure of conduction bands with DFT+U approach (U=2.0 eV) for (e) pristine NiPS\textsubscript{3}, (f) NiPS\textsubscript{3} monolayer with one K atom per unit cell, and (g) NiPS\textsubscript{3} monolayer with two K atoms per unit cell. (h) Core level spectra of the NiPS\textsubscript{3} under dosing measurements.}
\label{fig:APRES_dose}
\end{figure*}

\begin{figure*}[htb]
\centering
\includegraphics[width=7cm]{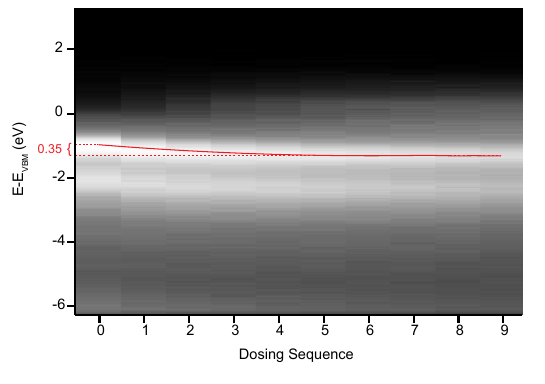}
\caption{\textbf{Band shifting with dosing.} As the dosing amount increases, the features of the valence band shift towards lower energies. The highest intensity band around $E-E_{VBM}$ = -1 eV decreases by approximately 0.35 eV.
}
\label{fig:shift}
\end{figure*}

Beyond the valence band electronic structure resolved by $\mu$-ARPES, the conduction band can also be probed via alkali metal doping (AMD), a technique that can efficiently add electrons to the system. The ARPES results for potassium (K) dosing are shown in Fig.~\ref{fig:APRES_dose}\textbf{a}-\textbf{c}. During the AMD process, charge transfers from K atoms to the surface of NiPS\textsubscript{3}, populating the formerly unoccupied conduction bands and shifting the entire band structure downward in energy. Over the course of dosing, the valence bands shift downward by approximately 0.35 eV (see Fig.~\ref{fig:shift}). The post-dosing spectra were energy-calibrated to the valence band features of the pristine sample, which remain largely unchanged aside from slight broadening due to increasing K coverage on the surface (see Fig. S2 in the Supplemental Material\cite{SM}). The upper panels of Fig.~\ref{fig:APRES_dose}\textbf{a}-\textbf{c} present the zoomed-in conduction band region. In comparison to pristine NiPS\textsubscript{3} (Fig.~\ref{fig:APRES_dose}\textbf{a}), two new bands at around 0.8 eV (red arrow) and 1.5 eV (orange arrow) emerge above the original $E_{VBM}$ in the doped samples (Fig.~\ref{fig:APRES_dose}\textbf{b}-\textbf{c}). As discussed in detail later, these correspond to a defect-related state and the low-lying conduction bands, respectively. The EDC in Fig.~\ref{fig:APRES_dose}\textbf{d} corroborate these observations: while the peaks of valence band remain at the same energy (albeit with a reduced intensity at higher K coverage), two new peaks appear at energies marked by red and orange arrows, corresponding to 0.8 eV and 1.5 eV, respectively.  Further analysis reveals that the 0.8 eV band is extremely flat and extends over the entire Brillouin zone, whereas the 1.5 eV band has a higher intensity and a broader bandwidth near the left $\overline{K}$ point, mirroring the similarly high-intensity valence band at -0.5 eV (marked by the uppermost white arrow in Fig.~\ref{fig:4}(a)) near that same $\overline{K}$ point.

To understand the newly observed ARPES features above \vbm, we perform DFT+U calculations using a 2×2 supercell of a monolayer NiPS\textsubscript{3} to model K adsorption at two distinct concentrations and compare the results with the pristine case (for details of the potassium doping calculation, see section 2.4 in the Supplemental Material\cite{SM}). Notably, the introduction of K atoms shifts the Fermi level of NiPS\textsubscript{3}, so to highlight features of valence bands, we align the VBM of the doped models with that of the pristine system, which is reasonable as indicated in the experimental results. The shifted VBM in Fig.~\ref{fig:APRES_dose}\textbf{e}-\textbf{g} is labeled as $E'$. In the pristine NiPS\textsubscript{3} model (Fig.~\ref{fig:APRES_dose}\textbf{e}), several conduction bands appear at around 1.5 eV above the VBM, matching the experimental onset of  conduction bands. When one K atom per unit cell is introduced, additional bands appear yet the general characteristics of both valence bands and conduction bands remain largely unaltered as shown in Fig.~\ref{fig:APRES_dose}\textbf{f}. Orbital analysis indicates that the lowest conduction band is primarily contributed by Ni and S orbitals, with no contribution from K atoms, implying that it is intrinsic to NiPS\textsubscript{3} rather than a doping-induced band (see Fig. S10 in the Supplemental Material\cite{SM}). Moreover, since both the experiment revealed lowest conduction band and the highest valence band are located at the same \emph{k} vector near the $\overline{K}$ point, NiPS\textsubscript{3} nanometer-thin flakes likely possess a direct band gap, in contrast to the indirect gap calculated for bulk NiPS\textsubscript{3} (Fig.~\ref{fig:1}\textbf{d}). Notably, the observed gap of 1.5 eV from ARPES lies remarkably close to the prominent spin-correlated PL peak at 1.476 eV\cite{kang2020coherent, wang2021spin, hwangbo2021highly}, suggesting that this PL feature indeed originates from transitions between these bands. Orbital-resolved calculations support this interpretation, revealing that the high-lying valence bands are dominated by Ni $d$ and S $p$ orbitals (Fig.~\ref{fig:4}\textbf{b} and \textbf{c}), whereas the low-lying conduction bands mostly consist of Ni $d$ orbitals in character (see Fig. S10 in the Supplemental Material\cite{SM}). Consequently, the spin-correlated PL corresponds to a predominantly local $d$-$d$ transition on Ni sites with partial S $p$ hybridization, in agreement with previous proposals\cite{kang2020coherent, klaproth2023origin, he2024magnetically}.

A remaining question is the origin of the flat band appearing around 0.8 eV. Unlike the low-lying conduction band at 1.5 eV, this band is highly localized and resides inside the band gap, hinting at defect states\cite{qi2018defect}. In ideal crystals, the periodic potential supports delocalized electronic states, but defects break this periodicity, creating localized potentials that introduce in-gap states. NiPS\textsubscript{3} can harbor natural or induced S or Ni vacancies\cite{tong2021dual, zhang2023boosted}, and related calculations suggest that S vacancies can produce defect levels\cite{wu2021adsorption}. The AMD process may increase defect formation on the NiPS\textsubscript{3} surface, explaining the emergence of defect state in the band gap. Even though more K doping will bring additional in-gap states above \( E_{VBM} \) as shown in Fig.~\ref{fig:APRES_dose}\textbf{g}, this two-K model disagrees with experiment in two ways: it dramatically alters the valence bands and splits the lowest conduction band into two sets (see Fig. S9 in the Supplemental Material\cite{SM}), whereas the new band does not split under further dosing as concluded from experimental observations (see Fig. S2 in the Supplemental Material\cite{SM}), nor does the K 3\textit{p} peak during the dosing process (Fig.~\ref{fig:APRES_dose}\textbf{h}). These inconsistencies imply that under our dosing conditions, the K concentration is below two atoms per NiPS\textsubscript{3} unit cell. Consequently, the flat band at 0.8 eV likely stems from defect states rather than a high doping level.

\begin{figure*}[htb]
\centering
\includegraphics[width=15.7cm]{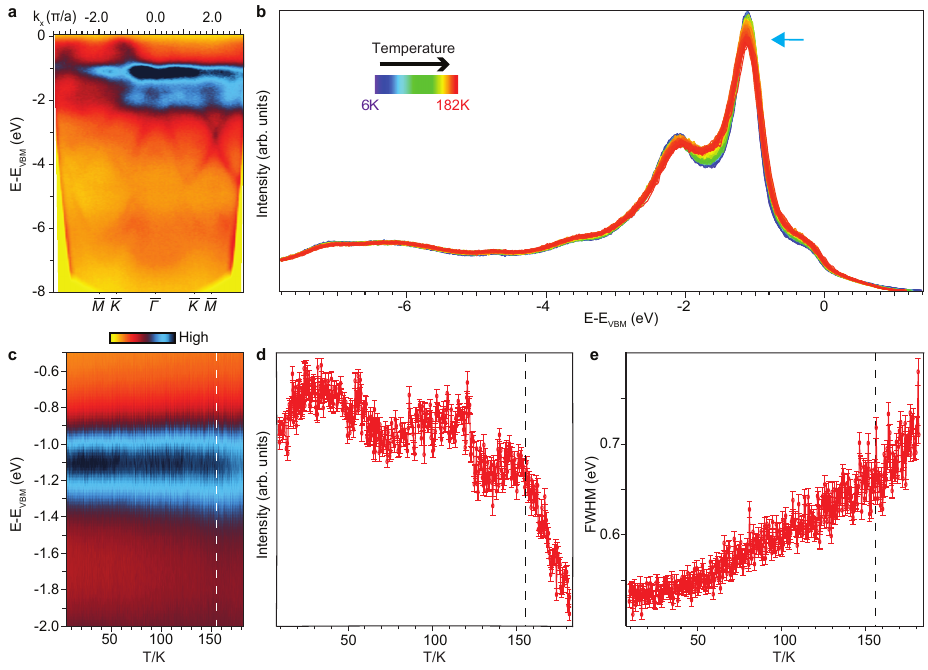}
\caption{\textbf{Temperature dependent band structure consistency.} (a) Experimental band dispersion of NiPS\textsubscript{3} in the paramagnetic state at 182 K. (b) Core level spectra of the NiPS\textsubscript{3} during temperature ramping from 6 K to 182 K with color bar indicating temperature change. (c)-(e) Temperature dependent behaviors of the strongest band at $\overline{\Gamma}$ point at around \evbm~= -1 eV indicated by blue arrow in (b) with (c) shape of band, (d) peak intensity, and (e) full width at half maximum (FWHM). The dashed lines represent the $T_N$ of \nps. }
  \label{fig:ARPES_tem}
\end{figure*}

All the experiments reported thus far have been conducted at a temperature of approximately 6 K, well within the AFM state of NiPS\textsubscript{3}. To explore whether the band structure of NiPS\textsubscript{3} changes above its $T_N$ of 155 K, we have performed measurements at various temperatures up to 182 K. Figure~\ref{fig:ARPES_tem}\textbf{a} shows the $\mu$-ARPES data of NiPS\textsubscript{3} in its paramagnetic state at 182 K. Although enhanced thermal motion blurs the bands slightly, and a slight upward shift occurs due to charging effect, the band structure at 182 K remains essentially the same as that at 6 K. Figure~\ref{fig:ARPES_tem}\textbf{b} presents the temperature-dependent density of states, revealing no significant changes in peak positions or intensities. Figures~\ref{fig:ARPES_tem}\textbf{c}–, which is measured at \(\overline{\Gamma}\) point, focus on the strongest peak at around -1 eV that is attributed to the flat bands arising from Ni \( d \) orbitals, as indicated by the blue arrows in Fig.~\ref{fig:ARPES_tem}\textbf{b}. Above $T_N$ (dashed line), the band position has basically no shift (Fig.~\ref{fig:ARPES_tem}\textbf{c}), yet the peak intensity shows a clear drop (Fig.~\ref{fig:ARPES_tem}\textbf{d}). In addition, the band width (FWHM) indicates no systematic trend changes across \( T_N \) (Fig.~\ref{fig:ARPES_tem}\textbf{e}). Overall, the ARPES spectra of NiPS\textsubscript{3} remain largely unaffected by crossing \( T_N \). This contrasts sharply with MnPS\textsubscript{3}, where Jeff \textit{et al.}\cite{strasdas2023electronic} reported pronounced splitting in the Mn \(3d\) bands near the \(\Gamma\) point and subtle changes in the S \(3p\) bands upon transitioning from the paramagnetic state to the AFM state. Such pronounced changes are likely suppressed in NiPS\textsubscript{3} because of the highly localized Ni \(3d\) states, as observed here and noted in earlier work\cite{kim2018charge}. DFT calculations further corroborate this view, showing that the local electronic structure of the Ni ions in NiPS\textsubscript{3} remains essentially independent of magnetic order\cite{chen2020paramagnetic}.

\section{Conclusions}

In summary, we have directly visualized the valence and conduction band structure of NiPS\textsubscript{3} using synchrotron-based $\mu$-ARPES combined with alkali metal dosing and density functional theory. The theoretical calculations facilitate the unambiguous identification of orbital contributions from Ni $d$, S $p$, and P $p$ states to the observed band structure. With potassium dosing, additional electronic states emerge: a conduction band at around 1.5 eV, alongside a defect-related state at around 0.8 eV. The measured band gap of around 1.5 eV closely matches the spin-correlated emission energy, strongly suggesting that this emission originates from a transition between the Ni $d$-S $p$ valence band and the Ni $d$ conduction band. Temperature-dependent ARPES measurements of NiPS\textsubscript{3} across $T_N$ reveal no notable changes in the band structure, in stark contrast to MnPS\textsubscript{3}, emphasizing the highly localized nature of Ni $d$ states. Overall, our findings may help deepen the understanding of the electronic landscape of NiPS\textsubscript{3} and provide essential insights for interpreting its intriguing physical properties. 

\section*{Data availability}

Data that support the study in this article are available from the corresponding authors upon reasonable request.

\section*{Acknowledgements}

This research used resources of the Advanced Light Source, which is a Department of Energy (DOE) Office of Science User Facility under contract no. DE-AC02-05CH11231. Work done by Q.T. and X.L.is supported by National Science Foundation (NSF) under Grant No. 1945364. Work done by H.G. and X.L. is supported by DOE, Office of Science Basic Energy Science (BES) under Award Number DE-SC0021064. X.L. acknowledges the membership of the Photonics Center at Boston University. M.J.S.M., C.G.V., and M.S.C.M. formally acknowledge the financial support provided by CNPq, CAPES, FAPEMIG, the Brazilian Institute of Science and Technology (INCT) in Carbon Nanomaterials, and Rede Mineira de Materiais 2D (FAPEMIG). The authors are also grateful to the National Laboratory for Scientific Computing (LNCC/MCTI, Brazil) in São Paulo (CENAPAD-SP) and the SDumont supercomputer in Rio de Janeiro for their invaluable assistance. Furthermore, M.J.S.M., C.G.V., and M.S.C.M. express appreciation for the postdoctoral scholarship funded by project CNPq-405910/2022-3. M.J.S.M additionally acknowledges the financial support extended by Universidade Federal de Ouro Preto (UFOP).

\section*{Author contributions}
Y.C., Q.T. and Y.G. contributed equally to this work. Y.C., Q.T., X.L. and K.E.S. conceived the project. Q.T. and H.G. synthesized, characterized, and prepared the NiPS\textsubscript{3} samples. Y.C., Q.T., Y.G., N.R., C.J., A.B. and E.R. conducted the synchrotron-based $\mu$-ARPES measurements. J.L. performed the lab-based ARPES measurements. M.J.S.M., C.G.V. and M.S.C.M. carried out the theoretical calculations. Y.C., Q.T., Y.G. performed the data analysis and interpretation with input from C.G.V. and M.J.S.M. Y.C., Q.T. and Y.G. wrote the manuscript with assistance from all authors.

\section*{Competing interests}
The authors declare no competing interests.

\bibliography{references}

\section{Figures}

\end{document}